\documentclass[conference]{IEEEtran}
\IEEEoverridecommandlockouts
\usepackage[hidelinks]{hyperref}
\usepackage{mathtools}
\usepackage{cite}
\usepackage{amsmath,amssymb,amsfonts}
\usepackage{algorithmic}
\usepackage{graphicx}
\usepackage{textcomp}
\usepackage{xcolor}
\usepackage{enumitem}
\usepackage{pgfplots}
\usepackage[hidelinks]{hyperref}
\usepackage{mathtools}
\usepackage{wrapfig}
\usepackage{siunitx}
\def\BibTeX{{\rm B\kern-.05em{\sc i\kern-.025em b}\kern-.08em
    T\kern-.1667em\lower.7ex\hbox{E}\kern-.125emX}}
\begin{document}
%
\title{Hierarchical Provision of Distribution Grid Flexibility with Online Feedback Optimization}

\author{
\IEEEauthorblockN{Florian Klein-Helmkamp, Irina Zettl,\\Florian Schmidtke and Andreas Ulbig}
\IEEEauthorblockA{IAEW at RWTH Aachen University\\
Aachen, Germany}
\and
\IEEEauthorblockN{Lukas Ortmann}
\IEEEauthorblockA{OST - Eastern Switzerland University of Applied Sciences \\
Rapperswil, Switzerland}
}

\maketitle

\begin{abstract}
Utilizing distribution grid flexibility for ancillary services requires the coordination and dispatch of requested active and reactive power to a large number of distributed energy resources in underlying grid layers. This paper presents an approach to hierarchically dispatch flexibility requests based on Online Feedback Optimization (OFO). We implement a framework of individual controllers coordinating actors, contributing to flexibility provision, to track a requested operating point at the interface between grid layers. The framework is evaluated in terms of performance during coordination and possible interaction between individual controllers, both central and distributed. Results show high reliability and robustness of the OFO controllers as well as an efficient dispatch of active and reactive power. Its computational efficiency and capabilities in set point tracking during online grid operation are making OFO a promising approach to the flexibility dispatch problem.
\end{abstract}


\section{Introduction}
\label{sec:Introduction}
\begin{figure}[tb]
    \centering
    \includegraphics{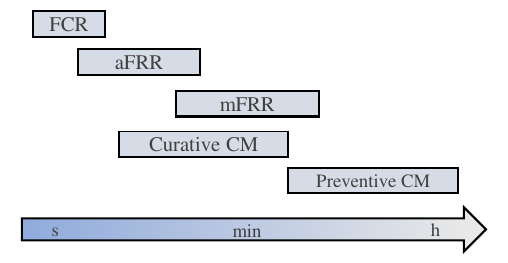}
    \caption{Time-scales of different ancillary services on transmission system level.}
    \label{fig:ancillaryservices}
\end{figure}

Utilizing flexibility from distribution grids is discussed as a possible measure for ancillary services, e.g. congestion management (CM) on transmission system level \cite{InnoSys, Silva_2018, Kolster_2020, Kolster_2022}, automatic and manual frequency restoration reserve \cite{Haberle_2022, Karagiannopoulos_2020} or voltage support \cite{Karagiannopoulos_2021}. Efficient and robust control of numerous distributed energy resources (DER) across multiple grid layers is necessary to leverage the existing flexibility in underlying distribution grids. This requires coordination methods designed to address the use-case specific challenges, that arise from the different types of ancillary services (see \autoref{fig:ancillaryservices}). These might include: Information asymmetry between grid operators, short response times for ancillary services or incomplete distribution grid models in operation \cite{Alazemi_2022}. For all types of ancillary services, these challenges have to be addressed by the implemented control architecture. Otherwise, the system might reach critical states, damaging equipment or negatively impacting system stability. A dispatch of DER in online grid operation must therefore be robust against disturbances and computationally efficient to ensure the provision of flexibility in an appropriate time frame. The general type of dispatch problem for DER is conventionally formulated as an optimal power flow (OPF) problem. While guaranteeing optimal results in the sense of the optimization problem, the disadvantages of OPF might prevent its usage for ancillary services with high requirements on robustness in coordination and response time. These include, i.a. proneness to sub-optimal results and no guarantee for constraint satisfaction in case of model mismatch, as well as high computational expenses due to its non-convex nature \cite{Molzahn_2017}. Online Feedback Optimization (OFO) has been proposed as a possible alternative, addressing the challenge of robustness by incorporating an optimization problem in closed loop with real-time measurements from the physical grid. This allows it to iteratively steer the set points for DER, tracking an optimal solution to the dispatch problem in online grid operation by solving the formulated problem on the physical system \cite{Picallo_2020}.

\subsection{Related Work}
\label{subsec:Related Work}
Incorporating distribution grid flexibility in online grid operation requires controlling the load flow across the point of common coupling (PCC) to the superimposed grid layer by dispatching new operating points to individual DER. Several approaches for the vertical coordination of flexibility in grid operation have been proposed so far. A formulation of a possible dispatch problem based on the second-order cone relaxation of the OPF is described in \cite{Torbaghan_2020}. Implementing a two-stage approach \cite{Kalantar_2021} is tackling the problem of flexibility coordination by utilizing a linear OPF formulation in real-time processes during grid operation. Quantifying the available flexibility potential, a method to describe feasible operating points for underlying distribution grids based on aggregating linear OPF models is proposed in \cite{Contreras_2021}. With the flexibility potential for distribution grids computed for a single operating point, the dispatch of requested set points is often done using feed-forward optimization. A proposed method based on a decoupled AC/DC OPF with feed-forward optimization is described in \cite{Olsen_2021}. This approach is lacking robustness in online grid operation and is highly dependent on the availability of accurate grid models. Especially for the case of distribution grids, complete information and models are not always available. Addressing this, a method for disaggregation of requested flexibility based on a linear OPF model is described in \cite{Frueh_2023}. It includes measuring the grid state, but is still reliant on an explicit model of the grid, as well as a state estimation within the control loop with the physical system.

To counteract the need for highly accurate models for online grid operation, OFO has been described in literature as a novel approach with a large number of possible application for the control of power systems. It is based on implementing an optimization algorithm in a closed loop with real-time measurements from the physical system \cite{Hauswirth_2021} and a linear approximation of the relation between system input and output \cite{Bolognani_2015}. A possible implementation for a controller solving a time-adaptive AC OPF using a mapping of input-output-sensitivities is proposed in \cite{Picallo_2022}. The presented approach holds several significant advantages over conventional feed-forward optimization methods, i.e. robustness in tracking the solution of the OPF and a low need for explicit model information about the controlled grid. Apart from techno-economic dispatch problems, several other applications for OFO in power system operation have been shown. An approach to control reactive power flow by OFO is presented in \cite{Ortmann_2020} with an experimental validation for the use-case of voltage control and a subsequent field test showing further applicability in \cite{Ortmann_2023}. The usage of OFO as a possible approach for real-time curative system operation was shown in \cite{Ortmann_2023_2}. An experimental validation of flexibility provision using OFO for a single grid layer in a laboratory setup is described in \cite{KH_2023}. The presented method is extended in this paper for a hierarchical setup over multiple grid layers.

\subsection{Main Contribution}
\label{subsec:Research}
In this paper, we present an approach to the hierarchical provision of distribution grid flexibility for ancillary services based on OFO. We propose an architecture of multiple OFO controllers acting in a hierarchy to adjust the flow of active power at the PCC between grid layers by actuating either DER connected to their respective system or further underlying OFO controllers. We define areas of controllability and observability for each OFO controller, as well as a hierarchy allowing superimposed controllers to request set points for active power from underlying grids. We thereby decouple the optimization problems for the individual grid layers, making them solvable on timescales that are relevant to real-time grid operation. This holds significant advantages over conventional OPF calculation for use-cases where flexibility has to be provided in a short interval of time, e.g. curative congestion management. We formulate the following two research questions:
\begin{enumerate}
    \item How can a hierarchical OFO control architecture be formulated for distributed provision of flexibility for ancillary services such as congestion management during online grid operation?
    \item Is interaction between the central OFO controllers and distributed volt/var controllers during flexibility coordination effecting the performance of the proposed architecture?
\end{enumerate}
Based on previous simulative studies and experimental validation we expect the hierarchical setup of controllers to provide accurate set-point tracking within the constraints of the individual grid layers. Furthermore, we expect the central operating controller to interact with the distributed volt/var controllers of each grid layer on a time-scale of several seconds to minutes while being able to ensure constraint satisfaction \cite{KH_2023}.

\section{Flexibility Provision}
\label{sec:Flex}
Flexibility provision from distribution grids is generally possible for a number of different ancillary services, that are subject to use-case-specific requirements and constraints. These might include the response time of the ancillary service or the robustness in coordination after the measure is activated. In this paper we focus on the provision of flexibility for preventive and curative congestion management on the transmission system level. This entails a time frame of several minutes with high demands on robustness to prevent permanent damage to temporarily overloaded operational resources, i.e. transmission lines or transformers \cite{KH_2023, Kollenda_2023}. From this we derive the basic requirements for the implemented controller architecture:
\begin{enumerate}
    \item High reliability and robustness in set point tracking.
    \item Fast and efficient provision of requested flexibility at the PCC.
    \item No permanent violations of operational limits of the flexibility providing grid.
\end{enumerate}
In the following subsections, we present the implemented controller hierarchy and evaluate it in terms of above the mentioned requirements to flexibility provision in two exemplary case studies.

\subsection{Hierarchical System Architecture}
\label{sec:Hier}
\begin{figure}[tb]
    \centering
    \includegraphics{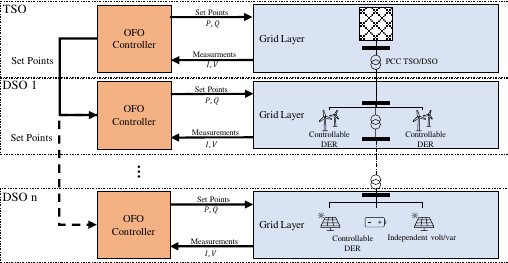}
    \caption{Hierarchy of individual OFO controllers}
    \label{fig:Hierarchy}
\end{figure}

In this section the hierarchical structure of OFO controllers proposed in this work is described in detail. The general structure of the hierarchy is oriented on the system boundaries, e.g. the transformers interfacing transmission and distribution grid. For the hierarchy we define a directed graph $G=(V,E)$ with the vertices $V$ corresponding to the actors requesting or providing flexibility in grid operation and the edges $E$ to the possible direction of flexibility requests. We assume the flexibility requests to always be downstream, i.e. from superimposed to underlying grid operators. Flexibility connected to the distribution grid is therefore activated with the goal of providing an ancillary service on the top-level grid layer. For each individual OFO controller we define a set of observable nodes $N$ and observable branches $B$ that correspond to the nodes and branches within its respective grid layer, as well as the PCC to the underlying grid layer. Furthermore, we define a set of controllable actors $F$ that are connected directly to the grid layer the corresponding controller is assigned to. This set includes all controllable DER and loads, as well as all PCC interfacing with underlying grid layers. Therefore, all underlying OFO controllers are part of the set of flexible actors $F$ of exactly one superimposed operator. All controllable actors can only be assigned to one OFO controller in the hierarchy to ensure that flexibility can only be directly actuated by the controller responsible for the grid layer they are connected to. The individual instances of OFO controllers are therefore acting as interfaces at the system boundaries. This has two main advantages from the perspective of a system operator. First, the area of observation and controllability is significantly smaller for each instance leading to a higher computational efficiency. Second, the amount of information to be shared between grid operators is kept at a minimum, counteracting possible privacy concerns regarding exchanged data. For flexibility provision the OFO controllers in the framework are iteratively measuring voltage and load flow at their nodes and branches and are sending set points for active and reactive power based on their internal optimization problem as shown in \autoref{fig:Hierarchy}. The underlying OFO controllers are thus updating their internal optimization problem with each time step they receive a new set-point for flow of active power at the PCC to the superimposed grid layer. The coordination of set points between individual operators in the hierarchy is therefore following the iterative nature of OFO. To quantify the available flexibility for the underlying grid layers, we perform an outer approximation of controllable active and reactive power for the current operating points of the DER connected to the underlying grid $F_{u}$. As the superimposed OFO controller is not directly responsible for the control of the requested set point at the PCC, the outer approximation is valid under the assumption that the underlying controllers are ensuring constraint satisfaction for their grid layer. An exemplary behavior of the hierarchy under active constraints is shown in \autoref{sec:Case Study}. The cycle time $\Delta t_{i}$ of the controller can be chosen freely for each OFO instance $i$ in the hierarchy and is therefore adjustable to the specific technical constraints of the individual grid layers, e.g. delays of control and measurement signals. Moreover, the setting of the cycle time can be used to implement the requirements for a specific ancillary service as defined in the respective grid codes. Within the hierarchical framework we define two general types of OFO controllers with different objectives in grid operation:
 \begin{enumerate}
     \item \textit{Primary Controller:} Supervision of the top-level grid layer.
     \item \textit{Secondary Controller:} Provision of flexibility to superimposed grid layer.
 \end{enumerate}
For a given scenario of operation across multiple grid layers, this classification results in a single primary OFO controller with one or multiple secondary OFO controllers that can themselves actuate OFO controllers of further underlying grid levels. For the architecture presented in this paper the optimization problem only differs for the controller on the top-level grid layer and all underlying instances. All controllers are measuring the state of their grid layer and are keeping bus voltages and branch flows within permissible ranges.

\subsection{Flexibility Dispatch Problem}
\label{subsec:Problem}
To efficiently provide operational flexibility we formulate two individual optimization problems that are integrated into the respective OFO controllers. The problems vary depending on the type of OFO controller in grid operation. For the primary OFO controller the optimization problem is chosen as:
\begin{equation} \label{eq:opti_prob_primary}
\begin{aligned}
    \min              &\quad&\Phi =  \sum_{i \in F} ||P_{\text{j}} - P_{\text{cur,j}}||^{2}  &&     &&     & 
    \\
    \text{s. t.}       &\quad&   V_{\text{min,n}}   &\leq V_{\text{n}} & \leq V_{\text{max,n}} &\quad \forall n \in N\\
                        &\quad&   S_{\text{min,i}} &\leq S_{\text{i}} &\leq S_{\text{max,i}} &\quad \forall i \in B\\
                        &\quad&   P_{\text{min,j}}   &\leq P_{\text{j}} & \leq P_{\text{max,j}} &\quad \forall j \in F\\
                        &\quad&   Q_{\text{min,j}}   &\leq Q_{\text{j}} & \leq Q_{\text{max,j}} &\quad \forall j \in F &
\end{aligned}
\end{equation}
The objective function is minimizing the curtailed active power on the highest grid level. The first two constraints are enforcing the operational limits of the gird, i.e. voltage band and line loading. By incorporating online measurements from the system, the power flow equations do not need to be included as equality constraints, which allows the usage of the simplified model presented in \autoref{eq:opti_prob_primary}. The third and fourth constraints are enforcing the feasible operational range for the different actors, that are connected directly to the top-level system and are able to provide flexibility for ancillary services. By constantly monitoring the bus voltages and load flows on the lines, the primary controller is able to detect violations of the permissible operating ranges in real time. It can react to them by utilizing flexibility from underlying grid layers, i.e. requesting set points for the power flow across the PCC from the secondary OFO controllers. The allocation and actuation of distribution level flexibility for congestion management is therefore done inherently in online grid operation. The requested flexibility is realized by the secondary controllers through adjustment of the operating points for the controllable DER or further underlying distribution grids connected in the flexibility providing grid layer. The optimization problem for every subsequent OFO controller is chosen as:
\begin{equation} \label{eq:opti_prob_secondary}
\begin{aligned}
    \min              &\quad&\Phi =  ||P_{\text{set}} - P_{\text{PCC}}||^{2}  &&     &&     & 
    \\
    \text{s. t.}       &\quad&   V_{\text{min,n}}   &\leq V_{\text{n}} & \leq V_{\text{max,n}} &\quad \forall n \in N\\
                        &\quad&  S_{\text{min,i}} & \leq S_{\text{i}} &\leq S_{\text{max,i}} &\quad \forall i \in B\\
                        &\quad&   P_{\text{min,j}}   &\leq P_{\text{j}} & \leq P_{\text{max,j}} &\quad \forall j \in F\\
                        &\quad&   Q_{\text{min,j}}   &\leq Q_{\text{j}} & \leq Q_{\text{max,j}} &\quad \forall j \in F &
\end{aligned}
\end{equation}
For the underlying OFO controllers the objective function is minimizing the difference between the measured flow of active power across the PCC with the superimposed grid layer and the received set-point. The constraints for the subsequent controllers are enforcing operational ranges for lines and buses, as well as for flexibility providing actors connected to the grid layer. Again, the power flow equations are not added explicitly to the model, because they are implicitly solved on the system to be controlled by operating in closed loop with the respective grid layer. By decoupling the individual optimization problems for each grid layer we ensure an efficient provision of flexibility while satisfying the constraints of the individual systems and allowing for different objective functions within the controllers.

\subsection{Online Feedback Optimization}
\label{subsec:OFO}
\begin{figure}[tb]
    \centering
    \includegraphics{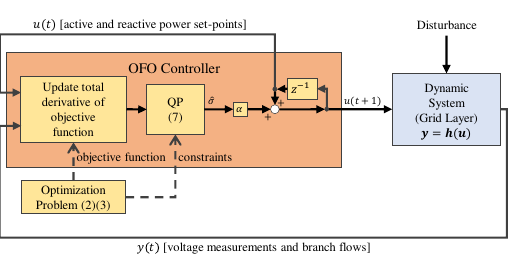}
    \caption{OFO controller for a single grid layer as implemented in this paper.}
    \label{fig:OFO_loop}
\end{figure}

\begin{figure}[b]
    \centering
    \includegraphics{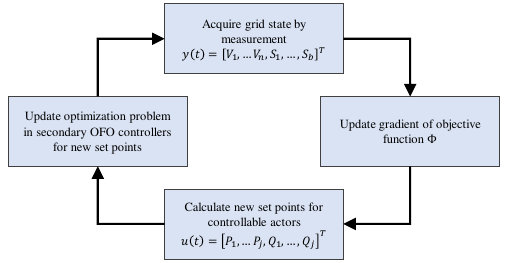}
    \caption{A single iteration of the OFO controller in grid operation.}
    \label{fig:OFO_cycle}
\end{figure}

To efficiently control requested distribution grid flexibility for ancillary services we implement the optimization problems as formulated in \autoref{subsec:Problem} in a closed loop with measurements from the respective grid layer. With $y(t)$ being a vector comprising measurements of bus voltage and branch flows, $u(t)$ being the vector of set points for active and reactive power and $\alpha \in \mathbb{R}^{>0}$ a positive scalar representing the fixed step-size parameter for the controller. For the given objective function $\Phi(u,y): \mathbb{R}^{p} \times \mathbb{R}^{n} \rightarrow \mathbb{R}$ the controller is iteratively finding the best constrained solution in online grid operation. First the current state of the system is acquired by measuring voltage and branch flow at relevant points in the grid for the current time-step $t$ with:
\begin{equation}
    y(t) = [V_{1}, ..., V_{n}, S_{1},...,S_{i}]^{T}
\end{equation}
Subsequently the gradient of the chosen objective function $\Phi$ is calculated for the current time-step $t$ with respect to the momentary values of the set points $u$ and measurements $y$ applying the chain rule as:
\begin{equation}
    \begin{aligned}
    \nabla \Phi(u,y)_{i} = &\nabla_{u} \Phi(u,y)|_{y=h(u)} \\
    &+ \nabla h(u)^{T} \nabla_{y}\Phi(u,y)|_{y=h(u)}
    \end{aligned}
\end{equation}
$\nabla h(u)$ is representing a steady-state input-output sensitivity matrix for each OFO controller in the hierarchy. It is determined a-priori for a given initial state of the respective distribution grid layer it is controlling with:
\begin{equation}
    \nabla h_{i,j} = \frac{\partial h(u)_{i}}{\partial u_{j}}
\end{equation}
As the OFO controller is implementing a gradient descent algorithm, the gradient of the cost function is updated for every iteration during flexibility provision. Following this, the step-size $\hat{\sigma}(u,y)$ is determined by solving an internal quadratic problem (QP), projecting the gradient of the objective function ${H}(u)^{T}\nabla \Phi(u,y)$ onto the set of feasible set points.
\begin{equation} \label{eq:ofostepcalculation}
	\begin{aligned}
		\hat{\sigma}(u,y) \coloneqq \arg \min_{w \in \mathbb{R}^{p}} \quad & \| w + H(u)^{T}\nabla \Phi(u,y)\|^{2}\\
		\textrm{s. t.} \quad & \begin{bmatrix}P_{min,j} \\Q_{min,j}\end{bmatrix}\leq \begin{bmatrix}P_j \\Q_j\end{bmatrix} + \alpha w \leq \begin{bmatrix}P_{max,j} \\Q_{max,j}\end{bmatrix}\\
		\quad & V_{min} \leq V_{meas} + \alpha \nabla h(u) w \leq V_{max} \\
  		\quad & S_{min} \leq S_{meas} + \alpha \nabla h(u) w \leq S_{max} \\\\
		\textrm{with} \quad &H(u)^{T} \coloneqq [ \mathbb{I}_{p} \hspace{2mm} \nabla h(u)^{T}]\\
        \textrm{and}  \quad & w \coloneqq \begin{bmatrix}
			                 \Delta P\\ \Delta Q
		                      \end{bmatrix}
\end{aligned}
\end{equation} 
Afterwards the vector of set points $u = [P_{1},...P_{j}, Q_{1},...,Q_{j}]^{T}$ for $t+1$ is calculated with:
\begin{equation}
    {u(t+1) = u(t) + \alpha\hat{\sigma}(u,y)}
\end{equation}
The new set points are then applied to the respective actors in the grid and the next iteration of OFO is performed after the pre-determined cycle time $\Delta t$. The sensitivity matrix $\nabla h(u)$ is representing the only explicit model information needed to implement the OFO controller for each grid layer. It expresses how a change in set point $u_{j}$ will affect the system output $y_{i} = h(u)_{i}$. Possible ways to determine the elements of $\nabla h(u)$ include exciting the system and measuring its response, data driven approaches as described in \cite{Picallo_2022} or offline calculation based on a system model. For the presented use-case $\nabla h(u)$ includes the sensitivities of a change in active and reactive power on the bus voltages as well as on the branch flows. The performance of the OFO controller can be tuned by adjusting the parameters mentioned above. A more in-depth exploration of the influence of individual parameters for the given use case is presented in \cite{Ortmann_2024}. Even though the sensitivity matrix is calculated for a fixed initial state of the dynamic system $y=h(u_{0})$, the algorithm itself is robust against possible model mismatch resulting from a different operating point by incorporating online measurements from the grid layer as shown in \autoref{fig:OFO_loop}. It is thereby holding several advantages over conventional feed-forward optimization approaches such as:
\begin{enumerate}
    \item Robustness in online grid operation against model mismatch and possible disturbances during flexibility coordination.
    \item Low need for explicit information, especially detailed models of the grid to be controlled.
    \item Increased computational efficiency compared to conventional OPF calculation.
\end{enumerate}
To enable the hierarchical provision of flexibility across multiple grid layers the requested set point for the flow of active power across the PCC is updated within the objective function in the optimization problem \eqref{eq:opti_prob_secondary} of the underlying controllers. A single cycle of the controller hierarchy as performed during flexibility coordination is shown in \autoref{fig:OFO_cycle}.

\section{Case Study}
\label{sec:Case Study}
In this section we showcase the proposed control architecture in three individual case studies. First, exemplary results for the control of a single grid layer are presented. This includes the investigation of the actuation of DER by the controllers and of the influence a flexibility provision has on the respective system. Second, the interaction between the central OFO controller and decentralized volt/var controllers during the coordination of flexibility is investigated for a single grid layer in an experimental setup on real hardware. Lastly, we present a simulative case study on a larger test system, comprised of a high, medium and low voltage layer.

\subsection{Single Layer Flexibility Dispatch}
\label{subsec:HierCoor}
\begin{figure}[tb]
    \centering
    \includegraphics{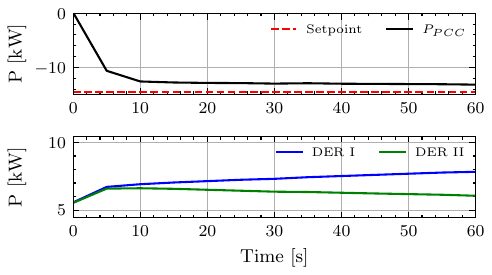}
    \caption{Converging behavior of controller for set point of $P_{\text{PCC}}=-14.5\unit{kW}$.}
    \label{fig:results_case_a}
\end{figure}

\begin{figure}[tb]
    \centering
    \includegraphics{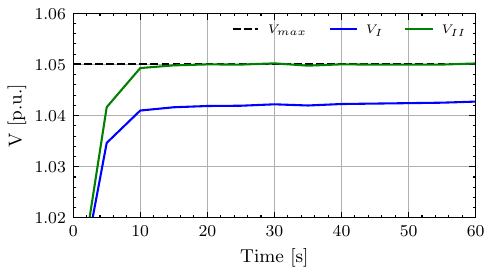}
    \caption{Voltage measurements of secondary OFO controller during flexibility provision.}
    \label{fig:results_case_a_volt}
\end{figure}

To quantify the performance of the presented controller hierarchy, we first look at its capability in reaching the requested set point for the flow of active power at a PCC by calculating the remaining deviation for $P_{\text{PCC}}$ after a defined interval of time $t$:
\begin{equation}
    \epsilon_{t} = \frac{|P_{\text{set}} - P_{\text{PCC}}(t)|}{|P_{\text{set}}|}
\end{equation}
A deviation might occur if a received set point is not fully reachable within the constraints of the flexibility providing grid for the defined time. The OFO controller is implemented for this case study on a single low voltage grid layer interfaced by a PCC to the superimposed medium voltage grid. This results in a single primary and single secondary OFO controllers in the hierarchy. For each of the OFO controllers we implement a cycle time of $\Delta t = 5s$ and a gain for the step-size of $\alpha = 0.22$. The optimization problem is chosen according to the position in the hierarchy (see \autoref{tab:case_1}).
\begin{table}[tb]
\begin{center}
\caption{Case Study A: Hierarchical setup of controllers}
\begin{tabular}{||c c c c c||} 
 \hline
 Instance & $\alpha$ & $\Delta t$ & Opt. Problem & Control Hierarchy\\ [1ex] 
 \hline\hline
 $OFO_{1}$ & 0.22 & 5 s & \eqref{eq:opti_prob_primary} & - \\ 
 \hline
$OFO_{2}$ & 0.22 & 5 s & \eqref{eq:opti_prob_secondary} & \{$OFO_{1}$, $OFO_{2}$\}\\ 
 \hline
\end{tabular}
\label{tab:case_1}
\end{center}
\end{table}
To showcase the controllers ability to provide a requested set point in grid operation, we first evaluate its performance for an exemplary request of ${P_{\text{PCC}} = -14.5 \unit{kW}}$. To reach the operating point, the OFO controller is iteratively adjusting the set points of the DER in its area of controllability (visible in \autoref{fig:results_case_a}). As this leads to a ramping up of active power, the bus voltages are consequently rising within the grid layer (see \autoref{fig:results_case_a_volt}). If a violation of the defined constraints (${V_{\text{N}} \pm 5\%}$) is detected, the OFO controller might not be able to fully realize the requested set point without actuating reactive power of the controllable inverters. As a result, a converging behavior to the set point can be observed, as shown in \autoref{fig:results_case_a}. For the presented case study, the controller is realizing the requested flexibility with an approximate error of ${\epsilon_{\text{60s}} = 7.3\%}$ after 60 seconds. As the voltage constraint is becoming active during flexibility provision, the secondary OFO controller is actuating the two DER differently to guarantee constraint satisfaction, exhibiting the reliability of the approach in grid operation. Through the iterative measurements of the branch flow at the PCC, the primary controller is able to react to set points that are not fully reachable by actuating other flexibilites in its controllability area. For the presented case study this leads to diverging set points for active power as visible in \autoref{fig:results_case_a}. With the bus voltage for DER II reaching the upper limit it is not further actuated to provide active power and DER I is dispatched to provide the remainder of requested flexibility within the constraints of the grid layer. As this behavior is strongly influenced by voltage control through reactive power, we present an experimental case study on the influence of reactive power provision, both centrally controlled and by independent volt/var control.

\subsection{Experimental Case Study with Independent Volt/Var Control}
\begin{figure}[tb]
    \centering
    \includegraphics{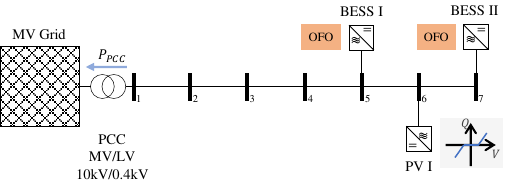}
    \caption{OFO controller for a single grid layer as implemented in this paper.}
    \label{fig:scenario_b}
\end{figure}

\begin{figure}[b]
    \centering
    \includegraphics{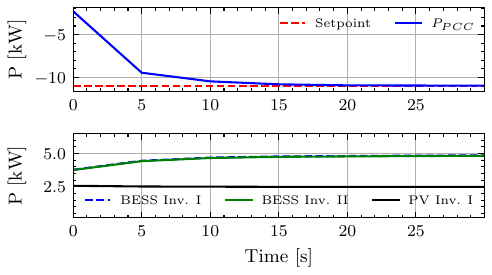}
    \caption{Set-point tracking by central OFO controller for a flexibility provision of $P_{PCC}=-11kW$.}
    \label{fig:results_set_point}
\end{figure}

\begin{figure}[t]
    \centering
    \includegraphics{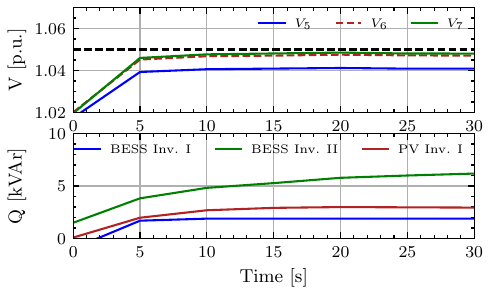}
    \caption{Interaction of central OFO controller and independent volt/var control during flexibility coordination.}
    \label{fig:results_volt_var}
\end{figure}

To investigate the interaction of the OFO controller with distributed volt/var control, we present an experimental case study for a single low voltage distribution grid with nominal voltage of $V_{N} = 0.4 \unit{kV}$ (see \autoref{fig:scenario_b}) in a laboratory setting. For the given scenario we implement a secondary OFO controller according to (\autoref{eq:opti_prob_secondary}) actuating two battery energy storage system (BESS) inverters. Additionally we include an independent photovoltaic (PV) inverter that is not controlled by OFO, but implements a volt/var control according to a droop curve if the bus voltage deviates more than $ \pm 3\%$ from the nominal voltage \cite{VDE_2018}. The technical parameters of the DER considered in the scenario can be seen in \autoref{tab:case_2_param}. The OFO controller in the case study is intentionally operating under limited observability to showcase the robustness of the approach. During coordination only bus voltages for the buses four, five, six, and seven are measured by OFO. The permissible voltage band for the central controller is set to $V_{N} \pm 5\%$. Branch flow, except for the active power at the PCC, is neglected. The secondary OFO controller is controlling the active power flow at the PCC to the superimposed medium voltage grid to a set point of $P_{PCC} = -11 kW$ at a cycle time of $\Delta t = 5s$ and with a gain of $\alpha = 0.22$. 
\begin{table}[tb]
\begin{center}
\caption{Case study B: Parameters of DER}
\begin{tabular}{||c c c c||} 
 \hline
 DER & $S_{rated}$ & Control Strategy &  Bus\\ [1ex] 
 \hline\hline
 PV I & 15 kVA & volt/var & 6 \\ 
 \hline
BESS I & 15 kVA & OFO & 5 \\ 
 \hline
 BESS II & 20 kVA & OFO & 7 \\
 \hline
\end{tabular}
\label{tab:case_2_param}
\end{center}
\end{table}
The results for a time-frame of 30s are shown in \autoref{fig:results_set_point}. Visible in red is the flow of active power at the PCC. The results show the OFO controller steering the operating point to the requested active power within three iterations and successfully tracking the given set point afterwards. The set point is reached by iteratively adjusting the in-feed for the two battery inverters (blue and green). Both BESS inverters are actuated identically to a set point of ${P_{\text{BESS I}} = P_{\text{BESS II}} = 4.8 \unit{kW}}$. The rest of the active power is provided by the independent PV inverter with a constant in-feed of $P_{\text{PV}}=2.5 \unit{kW}$. This results in an increase in bus voltage as shown in \autoref{fig:results_volt_var}. With the voltage rising above $V_{2}=1.03\;p.u.$ the volt/var control of the independent PV inverter is adjusting its in-feed of reactive power up to $Q_{\text{PV I}} = 3 \unit{kVAr}$, where it is kept for the remaining time. An interaction between the two control strategies is therefore visible in the case study. The independent volt/var controller is reducing the rise of bus voltage for its own bus and by extend for the neighboring buses. The OFO controller itself is only feeding-in reactive power if a constraint violation would occur without it for the application of the next set point vector $u(t+1)$. In consequence both BESS inverters are receiving set points for reactive power at $t=5s$. The secondary OFO controller is iteratively finding the optimal set points of reactive power to ensure constraint satisfaction while tracking the load flow at the PCC at the received set point. This results in stationary values of $Q_{\text{BESS I}} = 2 \unit{kVAr}$ and $Q_{\text{BESS II}} = 6 \unit{kVAr}$ and a total reactive power to enable the requested operating point for the distribution grid of $Q_{\text{total}} = 11.5\unit{kVAr}$. The results showcase the capabilities of the presented OFO controller in terms of constraint satisfaction during flexibility provision. By centrally controlling reactive power in addition to distributed volt/var control, the requested operating point is reachable within a few iterations and without permanent violation of operational constraints.

\subsection{Multi Layer Flexibility Dispatch}
\begin{figure}[tb]
    \centering
    \includegraphics{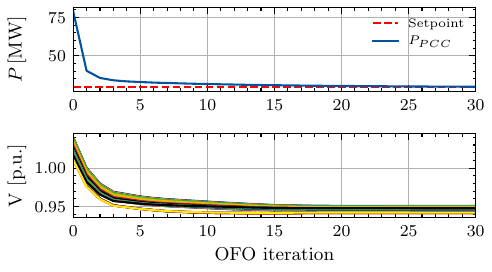}
    \caption{Adjustment of load flow across the PCC between HV and MV grid with resulting change in bus voltage in medium voltage grid.}
    \label{fig:case_c_p}
\end{figure}

\begin{figure}[tb]
    \centering
    \includegraphics{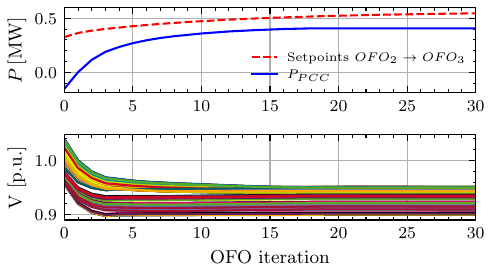}
    \caption{Interaction between $OFO_{2}$ and $OFO_{3}$ during flexibility coordination and bus voltages in flexibility providing low voltage grid.}
    \label{fig:case_c_setpoint}
\end{figure}

As a final case study we present the hierarchical interaction of multiple individual OFO controller on an expansive test system. The  scenario consists of multiple layers in the hierarchy of controllers. The investigated system is based on a \textit{Simbench} test case \cite{Simbench} and consists of an abstracted high voltage grid with a single PCC to the underlying medium voltage layer. Additionally we connect a low voltage grid to a medium voltage substation, offering flexibility that can be requested by the superimposed OFO controller. The hierarchy of OFO controllers and the corresponding optimization problems used in the control loops are shown in \autoref{tab:case_3}. We assume that all DER connected to both grid layers are controllable by the respective OFO controller. The limits for bus voltages vary between the grid layers with $0.95 \ \text{p.u.} \leq V \leq 1.05 \ \text{p.u.}$ for the medium voltage grid and $0.9 \ \text{p.u.} \leq V \leq 1.1 \ \text{p.u.}$ for the low voltage grid. The cycle time $\Delta t$ is neglected and the performance of the hierarchy is evaluated in terms of necessary iterations.
\begin{table}[tb]
\begin{center}
\caption{Case Study C: Hierarchical setup of controllers}
\begin{tabular}{||c c c c||} 
 \hline
 Instance & $\alpha$ & Opt. Problem & Control Hierarchy\\ [1ex] 
 \hline\hline
 $OFO_{1}$ & 0.5 & \eqref{eq:opti_prob_primary} & -\\ 
 \hline
$OFO_{2}$ & 0.05 & \eqref{eq:opti_prob_secondary} & \{$OFO_{1}$, $OFO_{2}$\}\\ 
 \hline
$OFO_{3}$ & 0.05 & \eqref{eq:opti_prob_secondary} & \{$OFO_{2}$, $OFO_{3}$\}\\
\hline
\end{tabular}
\label{tab:case_3}
\end{center}
\end{table}
To showcase the presented hierarchy of controllers, we simulate a set point of $P_{\text{set}} = 30 \unit{MW}$ for the flow of active power across the PCC of the high and medium voltage grid. The initial load flow in steady state at the PCC amounts to $P_{\text{PCC,0}} = 80 \unit{MW}$. The resulting adjustment to the flow of active power during coordination is shown in \autoref{fig:case_c_p}. It is visible how $90.2 \%$ of the requested set point at the PCC between high and medium voltage grid is fulfilled within 5 cycles of OFO with a remaining deviation of ${\epsilon_{\text{5}} = 9.8\%}$. After 25 cycles $P_{PCC} = 30 \unit{MW}$ is fully reached. During flexibility provision the bus voltages are dropping due to the curtailment of feed-in of DER in the system. As shown in \autoref{fig:case_c_p}, the voltage constraints can be satisfied during flexibility coordination. Each OFO controller is iteratively adjusting set points for active and reactive power of DER to ensure that bus voltages and line flows are kept within their respective limits. The interaction of the medium and low voltage layer in the given scenario is shown in \autoref{fig:case_c_setpoint}. The dashed line (red) is representing the set point that is requested by $OFO_{1}$ from $OFO_{2}$ in the current cycle of the control architecture. The underlying controller is therefore updating its objective function for each iteration with the new set point and is subsequently actuating DER connected to its area of controllability. The flow of active power across the PCC between the two grid layers is shown in blue. For the given scenario a converging behaviour is observable. This is due to the outer approximation of the operational range for the underlying low voltage grid. As $OFO_{2}$ is ensuring constraint satisfaction for the low voltage grid, the requested set point is not fully reachable. The superimposed $OFO_{1}$ is able to compensate for this deviation by actuating other DER in the medium voltage grid. The presented architecture is therefore exhibiting high robustness in case of incomplete or inaccurate operational ranges of flexibilities while still ensuring constraints for all participating grid layers during coordination.

\section{Conclusion}
\label{sec:Conclusion}
In this paper we presented an approach to coordinate and provide flexibility from DER for ancillary services by implementing individual OFO controllers in a hierarchy based on the interfaces between grid layers. Within the hierarchy each controller is assigned an area of controllability which corresponds to a single grid layer in flexibility provision or a subset of the connected DER. Requested operating points at the interface between individual grid layers are realized by iteratively adjusting the set points for active and reactive power in DER connected to the grid layer. We evaluated the performance of the hierarchical control structure in three individual case studies, both simulative and experimental on a test system in a distribution grid laboratory. For the given use-case and described implementation, OFO exhibits high reliability and accurate set point tracking during flexibility coordination by closing the loop through the incorporation of online measurements in each cycle. Furthermore, the low need for explicit model information and the reduction of computational effort due to the decoupling of individual grid layers are making the presented approach suitable for the flexibility dispatch problem across system boundaries. This holds especially true for the case of low voltage distribution grids, where accurate grid models and full observability is often not available. OFO is therefore a promising approach to leverage flexibility connected to underlying distribution grids for ancillary services with high requirements on robustness and reliability.

\end{document}